\documentclass[journal,10pt,final,twoside]{IEEEtran}
\usepackage[cmex10]{amsmath}
\usepackage{amssymb,amsfonts}
\usepackage{cite}
\usepackage{color}
\usepackage{graphicx,dblfloatfix}
\usepackage[tight,footnotesize]{subfigure}

\newtheorem{theorem}{Theorem}

\newtheorem{lemma}{Lemma}
\newtheorem{remark}{Remark}
\newtheorem{example}{Example}

\def\etal{{\em et al.}} 
\providecommand{\abs}[1]{\lvert#1\rvert} 
\newcommand{\set}[1]{\mathcal{#1}}
\newcommand{\indicator}{\mathbf{1}}
\def\sinr{\ensuremath{\mathsf{SINR}}}
\def\snr{\ensuremath{\mathsf{SNR}}}
\newcommand{\PP}{\Pr}    
\newcommand{\Prob}[1]{\PP [#1]}
\newcommand{\EE}{\mathbb E}                    
\newcommand{\expt}[1]{\EE [#1]}
\newcommand{\ud}{\mathrm{d}}

\newcounter{MYtempeqncnt}

\begin{document}

\title{Throughput Scaling of Wireless Networks\\
With Random Connections}
\author{Shengshan~Cui,~\IEEEmembership{Member,~IEEE,}
Alexander~M.~Haimovich,~\IEEEmembership{Senior~Member,~IEEE,} \\
Oren Somekh,~\IEEEmembership{Member,~IEEE,} H.~Vincent~Poor,~%
\IEEEmembership{Fellow, IEEE,} and Shlomo~Shamai~(Shitz),~%
\IEEEmembership{Fellow, IEEE}\thanks{%
Manuscript submitted September 23, 2008; revised October 24, 2009.}\thanks{%
The work of S.~Cui and A.~M.~Haimovich is supported in part by the
National Science Foundation under Grant CNS-0626611. The work of
O.~Somekh is supported in part by a Marie Curie Outgoing
International Fellowship within the 7th European Commission
Framework Programme. The work of H.~V.~Poor is supported in part by
National Science Foundation under Grant CNS-09-05086 and in part by the Air Force Office of Scientific Research under Grant FA9550-08-1-0480. The work of S. Shamai is supported by European
Commission in the framework FP7 Network of Excellence in Wireless
COMmunications NEWCOM++. The material in this paper was presented in part at the IEEE International Conference on Communications, Dresden, Germany, June 2009.}
\thanks{%
S. Cui was with the Department of Electrical and Computer Engineering,
New Jersey Institute of Technology, Newark, NJ 07102 USA. He is now with
Qualcomm Inc., San Diego, CA 92121 USA (e-mail: scui@qualcomm.com).}
\thanks{%
A.~M.~Haimovich is with the Department of Electrical and
Computer Engineering, New Jersey Institute of Technology, Newark, NJ
07102 USA (e-mail: alexander.m.haimovich@njit.edu).}
\thanks{%
O.~Somekh is with Yahoo! Labs Israel, and with the Department of Electrical Engineering, Princeton University, Princeton, NJ 08544, USA (e-mail: orens@princeton.edu).}
\thanks{%
H.~V.~Poor is with the Department of Electrical Engineering,
Princeton University, Princeton, NJ 08544, USA (e-mail: poor@princeton.edu).}
\thanks{%
S. Shamai (Shitz) is with the Department of Electrical Engineering,
Technion--Israel Institute of Technology, Technion City, 32000 Haifa, Israel
(e-mail: sshlomo@ee.technion.ac.il).}}%

\maketitle

\newpage
\begin{abstract}\boldmath
This work studies the throughput scaling laws of ad hoc wireless
networks in the limit of a large number of nodes. A random
connections model is assumed in which the channel connections
between the nodes are drawn independently from a common
distribution. Transmitting nodes are subject to an on-off strategy,
and receiving nodes employ conventional single-user decoding. The
following results are proven:

\begin{enumerate}
\item For a class of connection models with finite means and variances, the
throughput scaling is upper-bounded by $O(n^{1/3})$ for single-hop
schemes, and $O(n^{1/2})$ for two-hop (and multihop) schemes.

\item The $\Theta (n^{1/2})$ throughput scaling is achievable for a specific
connection model by a two-hop opportunistic relaying scheme, which
employs full, but only local channel state information (CSI) at the
receivers, and partial CSI at the transmitters.

\item By relaxing the constraints of finite mean and variance of the
connection model, linear throughput scaling $\Theta (n)$ is
achievable with Pareto-type fading models.
\end{enumerate}
\end{abstract}


\section{Introduction}
\label{sec:intro} \PARstart{C}{onsider} a communication network in
which $n$ source nodes transmit data to their designated destination
nodes through a shared medium, with or without the help of
intermediate relays. Possible communication protocols include
single-hop schemes, where sources communicate directly with their
destinations without the help of relay nodes, two-hop schemes, where
sources communicate with destinations via intermediate relay nodes
in a two-hop fashion, and multihop schemes, where
source-to-destination (S--D) communication is done through multiple
layers of relays.

Two fundamental questions of interest for such networks are as
follows:

\begin{enumerate}
\item What is an efficient communication protocol for operating the
wireless network?

\item Given a specific protocol, what is the throughput performance, and
how one can design an operational scheme to approach the optimal
performance?
\end{enumerate}

These fundamental problems have led to an explosion of research
activity over the last decade. In particular, inspired by Gupta and
Kumar \cite{GK:00}, there has been significant recent progress in
understanding the \emph{throughput scaling of wireless ad hoc
networks}, i.e., the throughput performance in the large
system limit. Depending on the network model and on the level of channel
state information (CSI) available at the nodes, a variety of results
have been published, e.g.,
\cite{GK:00,FDTT:07,GT:02,DGT:05,XK:04,XK:06,JVK:04,XXK:05,LT:05,OLT:07,GHH:06,AS:07,HN:05,CJ:08,DH:06,MB:07,EMK:07,CHSP:07,Etkin:06}.

\subsection{Background and Motivation}
%

In \cite{GK:00}, Gupta and Kumar considered a model in which $n$
S--D pairs are located in a unit area. They showed that in the
limit, with an increasing number of nodes $n$, classical multihop
architectures with conventional single-user decoding and forwarding
of packets can achieve an aggregate throughput scaling no higher
than $\Theta (\sqrt{n})$,\footnote{Throughout the paper, a natural
base logarithm is used. For two functions $f(n)$ and $g(n)$, the
notation $f(n)=O(g(n))$ means that $\abs{f(n)/g(n)}$ remains bounded
as $n\rightarrow \infty $. We write $f(n)=\Theta (g(n))$ to denote
that $f(n)=O(g(n))$ and $g(n)=O(f(n))$, and we write $f(n)=o(g(n))$
to denote that $\abs{f(n)/g(n)}\rightarrow 0$ as $n\rightarrow
\infty $. In addition, we write $f(n)\sim g(n)$ to denote
$\lim_{n\to\infty}{f(n)}/{g(n)}=1$. The indicator function is
denoted by $\mathbf{1} (\cdot )$.} even for nodes with arbitrary locations.
Furthermore, they showed that $\Theta (\sqrt{n/\log n})$ is
achievable with randomly located nodes and multihopping with nearest
neighbor communication. The gap in performance between
arbitrary and random nodes was later closed by Franceschetti \etal\
\cite{FDTT:07}. Grossglauser and Tse \cite{GT:02} considered a
mobility model, in which the nodes are assumed to move across the
area in an independent, stationary, and ergodic fashion, and they
showed that, in sharp contrast to the fixed node case, the $\Theta
(1)$ throughput per S--D pair is indeed achievable, implying a
network aggregate throughput of $\Theta (n)$. Follow-up work
\cite{DGT:05} showed that $\Theta(n)$ throughput scaling is also achievable
 with restricted mobility patterns.

The factor limiting the performance of the multihop scheme of
\cite{GK:00} is fundamentally the interference. Specifically,
single-user decoding at each receiver cannot mitigate the
interference stemming from concurrent transmissions. Thus, it is
important for the network to control the total generated
interference by prohibiting those long-range, direct communications
transmissions that require large power. Consequently, most
communications occur between nearest neighbors, which limits the
actual useful throughput per S--D pair, since much of the traffic
carried by the nodes must necessarily be relayed traffic. Loosely
speaking, with $n$ nodes randomly populating a unit area, the
typical distance for an S--D pair is $\Theta (1)$ and the distance
between typical nearest neighbors is $\Theta (1/\sqrt{n})$. This
implies that the packets have to be relayed $\Theta (\sqrt{n})$
times to reach the destination. In the mobility model of
Grossglauser and Tse \cite{GT:02}, communications are also confined
to nearest neighbors, but the scheme ingeniously gets around the
burden of relaying by allowing the relay nodes to hold the packets
until they meet the desired destination nodes. This essentially
entails a two-hop scheme that improves the throughput of fixed
networks \cite{GK:00} by the order of $\Theta (\sqrt{n})$, yielding
an
aggregate throughput of $\Theta (n)$. 

By recognizing the interference-limited nature of the Gupta-Kumar
setup \cite{GK:00}, there has been another line of work focusing on
implicit interference management. 
It turns out that the ability of handling interference management highly
depends on the CSI knowledge
\cite{OLT:07,CJ:08,GHH:06,AS:07,HN:05,DH:06,MB:07,EMK:07,CHSP:07}.

On one extreme of the CSI assumption, global CSI is assumed in
\cite{OLT:07,CJ:08,GHH:06,AS:07,HN:05}. By global CSI it meant that
the channel realizations are known \emph{a priori} to all the nodes.
It turns out that with global CSI, one can indeed achieve $\Theta
(n)$ throughput scaling. {\"{O}}zg{\"{u}}r \etal\ \cite{OLT:07}
considered a setup in which $n$ S--D links are randomly located in
an area, and the channel connection between any two nodes is subject
to a path-loss attenuation and random phase. They showed that a
hierarchical cooperation scheme operating within a dense network,
supports a total throughput that scales linearly with $n$. A key
ingredient of the hierarchical scheme is a point-to-point MIMO (multiple-input
multiple-output) setup applied in clusters of nodes that form distributed MIMO\ systems. In
doing so, interference due to concurrent transmissions is canceled
by applying receive MIMO techniques \cite{Foschini:96,telatar:99}.
Cadambe and Jafar \cite{CJ:08}
considered an $n$ user interference channel, and they showed that in
random connection models, where the channel gains are randomly drawn
form a continuous distribution, a total number of $n/2$ spatial
degrees-of-freedom per orthogonal time and frequency dimension can
be achieved in an almost surely sense. This result can be achieved
by interference alignment at the transmitter coupled with zero
forcing detection at the receiver. In interference alignment, all
transmitters align their signal vectors in such a manner that they
cast overlapping shadows at the receivers where they constitute
interference, while they remain distinguishable at the intended
receivers.

A less demanding CSI assumption was considered by \cite{DH:06} and
\cite{MB:07}. In these two-hop amplify-and-forward schemes, each
relay node is assumed to know its own channel connections of either
all source/destination nodes \cite{DH:06} or only a subset of
source/destination nodes \cite{MB:07}, but has no knowledge of the
channel realizations pertinent to the other relays. With $n$ source
nodes transmitting simultaneously, relays perform decentralized
maximum ratio combining across the backward (sources to relays) and
forward (relays to destination) channels, respectively. Here,
without global CSI, it is not possible to fully \emph{cancel}
interference. Instead, the schemes rely on the power gain due to the
maximum ratio combining (a form of coherent matched filter) to boost
the strength of the desired signal, and use this power gain to
\emph{compensate} for the interference power. With the assumption
that each relay node has full knowledge of its local channels
(backward channels from all source nodes, and forward channels to
all destination nodes), the scheme by Dana and Hassibi \cite{DH:06},
achieves a total throughput of $\Theta (n^{1/2})$. In contrast, with
the Morgenshtern--B{\"{o}}lcskei scheme \cite{MB:07}, a reduced CSI
requirement yields a reduced throughput of $\Theta (n^{1/3})$.

Continuing to explore schemes with reduced CSI requirements, a
somewhat conservative, yet practically interesting scenario is one
in which each receiving node in the network has perfect knowledge of
the \emph{local} channel realizations, but transmitting nodes have
only some partial form of CSI. In \cite{EMK:07}, Ebrahimi \etal\
investigated the throughput of single-hop wireless networks in a
Rayleigh fading environment. With an on-off transmitting strategy,
in which source nodes either transmit or keep silent under the
control of the destinations, the authors showed that the maximum
throughput scales as $\log n$ where $n$ is the total number of S--D
pairs. Furthermore, a decentralized scheme that achieves the optimal
scaling laws was proposed in \cite{EMK:07}. The scheduling policy of
the scheme is based on the SNR (signal-to-noise ratio) measured at
the receiver, and an on-off command fed back to the transmitters,
i.e., each destination node instructs its source node to transmit if
the SNR exceeds a predefined threshold. To enable SNR-based
scheduling, a low-rate feedback link is needed. In \cite{CHSP:07},
Cui \etal\ considered a two-hop relaying scheme and they showed
that in a Rayleigh fading environment, the optimal throughput scaling of the
two-hop setup is also $\Theta (\log n)$. Furthermore, an opportunistic
relaying scheme was proposed in \cite{CHSP:07} that achieves the $\Theta
(\log n)$ throughput scaling. In essence, the opportunistic relaying scheme
exploits the innate power gain due to the presence of multiple nodes in the
network, i.e., the so-called \emph{multiuser diversity} \cite{VTL:02}, to
overcome the interference. To schedule the source nodes favored by multiuser
diversity, low-rate feedback communication is required from receivers to
transmitters.

The works in
\cite{OLT:07,CJ:08,GHH:06,AS:07,HN:05,DH:06,MB:07,EMK:07,CHSP:07}
have clearly exemplified the impact of the availability of CSI on
the throughput performance. 
While the global CSI assumption might be feasible in certain scenarios, e.g.,
point-to-point MIMO systems where antennas are co-located, it poses a difficult challenge for ad
hoc networks, where nodes are distributed. Moreover, the
size of the ad hoc network may also prohibit the exchange of CSI
across the nodes. Therefore, it is of importance to
understand the performance limits of wireless ad hoc networks
operating with more practical CSI assumptions. This is the main goal
of this work.

\subsection{Main Results}

In this work, we address the questions raised at the outset of this section
under the following assumptions:

\begin{itemize}
\item \emph{Random connection model:} The channel connections between nodes are
assumed to be drawn independently from a common probability
distribution. We allow the latter to vary with the number of S--D pairs $n$.

\item \emph{CSI:} Throughout the work, it is
assumed that, at each hop, each receiver node has perfect CSI knowledge of
all transmitters, but transmitter nodes have only partial CSI via receiver
feedback.

\item \emph{Transmitting strategy:} The transmitters apply an on-off
transmission strategy, i.e., they either transmit at full power or keep
silent, depending on the receivers' feedback information.

\item \emph{Single-user decoding:} The receiving nodes apply conventional
single-user decoding, i.e., multiple access interference is treated as noise.
\end{itemize}

In the first part of the paper, we restrict ourselves to a class of
connection models with finite mean and variance, which encompasses a
wide range of well-accepted channel models encountered in wireless
communications. The results are summarized in two theorems.

\smallskip

\begin{theorem}[Main Result]
Under the aforementioned assumptions, for a class of fading
distributions with finite mean and variance, the throughput scaling
is upper-bounded by $O(n^{1/3})$ for single-hop schemes, and
$O(n^{1/2})$ for two-hop schemes. Furthermore, multihop schemes
cannot achieve better scaling than two-hop schemes.
\end{theorem}

\smallskip

\begin{theorem}
Under the same assumptions as those of Theorem~1, the $O(n^{1/2})$ average throughput
scaling upper bound for two-hop schemes is achievable by the two-hop
opportunistic relaying scheme under the random connection model
whose channel (power) connections $\gamma$ has following cumulative
distribution function (cdf):
\begin{gather}
F_{n}(x)=\biggl(\frac{(x-\mu )/\sigma }{\sqrt{2n-1}}\frac{n-1}{n}+%
\frac{1}{n}\biggr)^{1/(n-1)},  \label{eq:mysterious_fading} \\
\mu -\tfrac{\sqrt{2n-1}}{n-1}\sigma \leq x\leq \mu
+\sqrt{2n-1}\sigma , \notag
\end{gather}%
where $\mu $ and $\sigma^2$ are the finite mean and variance
of the channel connection, respectively.
\end{theorem}

\smallskip \smallskip

It is shown in Theorem~\ref{thm:upper_bound} that, for any realization of the fading process, the throughput is limited by interference. The theorem establishes the fundamental throughput limits of various communication protocols.  From the architecture's perspective, the theorem suggests the optimality of two-hop schemes
since they have better thoughput upper bound than single-hop
schemes. Moreover, the theorem implies that, under the random
connection model, there is no advantage to multihop schemes. As
will be apparent in Section~\ref{sec:proof_upper_bound}, a two-hop
scheme is able to extract all the freedom of routing the packets,
and multihop schemes cannot do any better.

Theorem~\ref{thm:opportunistic_by_MD} focuses on two-hop schemes and
augments Theorem~\ref{thm:upper_bound} by showing that $\Theta
(n^{1/2})$ average throughput (by averaging over fading processes)
scaling is indeed achievable by two-hop
schemes. This is established by a constructive example.
Specifically, it is shown that the two-hop opportunistic relaying
scheme, originally studied in \cite{CHSP:07} under Rayleigh fading channel,
achieves the throughput upper
bound $O(n^{1/2})$ under model \eqref{eq:mysterious_fading}. 
The significance of Theorem~\ref{thm:opportunistic_by_MD} is
two-fold: First, the achievability of $\Theta (n^{1/2})$ confirms
that the upper bound in Theorem~\ref{thm:upper_bound} is tight.
Second, the achievability of $\Theta (n^{1/2})$ by two-hop schemes,
and the $O(n^{1/3})$ throughput upper bound for single-hop schemes
(cf.~Theorem~\ref{thm:upper_bound}) tell us that two-hop schemes are
indeed superior to single-hop schemes from the throughput
standpoint. It is noted that the model of
\eqref{eq:mysterious_fading} entails a channel distribution
dependent on $n$, the number of nodes in the wireless network. Such
a distribution is not likely to be encountered in practice, hence
the result of Theorem~2, should be viewed as theoretical.
Nevertheless, it is also noted that the proof of Theorem~1 on the
throughput scaling $O(n^{1/2})$ upper bound, is general for channels
with finite mean and variance, and is not in any way restricted to
channels that depend on the number of users in the network.

\smallskip In the second part of the paper, we relax the restriction of
finite mean and variance of the underlying channels, and the goal is to see
whether linear throughput scaling for two-hop schemes is achievable with
this relaxation. We prove the following affirmative answer:

\smallskip

\begin{theorem}
Linear average throughput scaling of $\Theta (n)$ is achievable by a two-hop
relaying scheme under a class of fading models whose cdf $F$
satisfies
\begin{equation}
1-F(x)\sim \frac{c_0\cdot x^{-\nu}}{\Gamma (1-\nu)}
\label{eq:non_mysterious_fading}
\end{equation}%
where $0<\nu <1$, $c_0$ is a constant, and $\Gamma (x)$ is the gamma
function.
\end{theorem}

\smallskip

As an illustrative example of this theorem's application, we show
that the Grossglauser--Tse scheme \cite{GT:02} can be casted in the
framework of a random fading network, and consequently the linear
throughput $\Theta (n)$ of the Grossglauser--Tse scheme follows in
straightforward fashion from the theorem.

It is worth pointing out that the random connection models
\eqref{eq:mysterious_fading} and \eqref{eq:non_mysterious_fading}
(used in Theorem~2 and Theorem~3, respectively) serve to give
examples of achieving throughput limits; however, they do not
preclude the existence of other models from yielding the same
throughput scalings.

In summary, in this work, we consider the fundamental throughput
limits of wireless ad hoc networks under fading channels by imposing
assumptions that are implementable with the state-of-the-art technologies.
We are able to quantify
different schemes from an architectural perspective
(cf.~Theorem~\ref{thm:upper_bound}). We also give a complete picture
of the throughput scaling laws of wireless ad hoc network under
different fading channels
(cf.~Theorem~\ref{thm:upper_bound}--\ref{thm:linear_achiving_Pareto}).
Furthermore, specific schemes with decentralized operation and
pragmatic CSI assumptions are shown to achieve the throughput
scaling upper bound.

\subsection{Related Work}

Throughout this work, we adopt the broad perspective of the impact
of the channel connection models on the network throughput scaling.
Specifically, the discussion is based on a general assumption about
the connection models (cf. Theorem~\ref{thm:upper_bound}) instead of
particularizing to any specific model. Such specific analysis for
Rayleigh fading model, for example, can be found in \cite{Etkin:06}
and \cite{EMK:07} for single-hop setups, and
in \cite{CHSP:07} for a two-hop setup. 
Recently, other specific channel models have been studied. See, for
example, \cite{EMK:06} for a single-hop scheme in lognormal fading.

The random connection model of this work is set in a \emph{dense}
network in that, as $n\rightarrow \infty $, the channel connection
between any two nodes does not diminish. In the context of the
geometric model, a dense network is interpreted as the total
geographical area is fixed and the density of nodes increases as $n$
grows. Under the geometric model, the throughput scaling for
single-hop schemes had been identified by \cite[Th.~III-3]{GT:02},
with the main result being that the system throughput is limited by
interference, and the network throughout is upper-bounded by
$O(n^{1/(1+2/\alpha )}),$ where $\alpha $ is the path-loss exponent.

\subsection{Organization of the Paper}

The rest of the paper is organized as follows. In
Section~\ref{sec:sysmod}, the channel model and system model are
formally introduced. In Section~\ref{sec:proof_upper_bound}, we
discuss the throughput upper bound scaling under a class of fading
models possessing finite mean and variance.
Section~\ref{sec:achieve_square_root_n} continues the discussion
along this class of fading models, and proves the achievability of
$\Theta (n^{1/2})$ average throughput for two-hop schemes. In
Section~\ref{sec:beyond_squre_root_n}, we relax the restriction on
channel fading models and prove the achievability of a linear average
throughput scaling under a Pareto-type fading model. Finally, in
Section~\ref{sec:conclusion}, conclusions are drawn. Only the
essential formulae appear in the main text; detailed mathematical
derivations are placed in appendices.

\section{System Model}

\label{sec:sysmod}

Consider a dense network consisting of $n$ source nodes seeking to
communicate with $n$ destinations, with or without the help of
intermediate relay nodes. Specifically, the communication between
sources and destinations can take place in a single-hop fashion,
where sources communicate with destinations directly, in a two-hop
fashion, where the sources first communicate with relays nodes and
then the relay nodes forward the packets to destinations, or in a
multihop scheme, where S--D communication is facilitated by multiple
layers of relaying. We assume that in two-hop and multihop protocols,
for each layer of relaying, there are $m$ relays. Furthermore, we assume that relays have the capability to buffer data.

The channel and signal models presented below apply to all
aforementioned three protocols. We adopt the random connection model
of \cite{GHH:06}, where the channel gains between nodes are subject
to flat fading whose powers are independent and identically
distributed (i.i.d.) according to a common, but unspecified
distribution. Mathematically, the channel (power) connection
$\gamma_{i,j}$ between node $i$ and $j$ are i.i.d. random variables
with distribution $f(\gamma)$. Similar to \cite{GHH:06}, for maximum
generality, we allow $f(\gamma)=f_n(\gamma)$ to be a function of the
number of nodes $n$.\footnote{Throughout the paper, we shall
explicitly use subscript $n$ to denote the dependence of a function
on $n$.} In Section~\ref{sec:proof_upper_bound} and
\ref{sec:achieve_square_root_n}, we restrict the discussion to the
case where the channel distribution has finite mean and variance
(i.e., $\mathbb{E}[ \gamma] =\mu $ and $\mathrm{Var}[ \gamma]
=\sigma ^{2}$ are finite); we later relax this restriction in
Section~\ref{sec:beyond_squre_root_n}. Furthermore, a quasi-static
model is assumed, i.e., channel realizations are fixed at a given
hop, but change values independently between hops. Except in Section~\ref{sec:proof_upper_bound} in which we are allowed to relax the CSI assumption in establishing the throughput upper bound, we assume that, in each hop, only receivers are assumed to
have current, perfect knowledge of the backward channel
realizations, but transmitters have only partial CSI. More
specifically, each transmitter knows the state of transmission
(i.e., to transmit or not) via receivers feedback. Note that the CSI
assumptions are quite relaxed and pragmatic. For example, in a
wireless system, the CSI at the receiver can be learned through the
transmission of training signals, while partial CSI at the
transmitter can be obtained with the help of a low-rate feedback
from receivers to transmitters.

In this paper, we assume that communication through the relays is
accomplished by \emph{decode-and-forward}. In each hop, transmitters
apply an \emph{on-off strategy} (depending on the feedback
information) and receivers apply \emph{conventional single-user
decoding}. More specifically, in Section~\ref{sec:proof_upper_bound} in which the throughput upper bound is studied, we assume genie-aided scheduling, i.e., a genie is assumed to have global CSI knowledge of the network. The genie feeds back information to control transmissions from a subset of the source nodes such that throughput is maximized. In Section~\ref{sec:achieve_square_root_n} and \ref{sec:beyond_squre_root_n}, in which a specific scheme is studied, we assume that CSI is available only at the receivers, and that the receivers feed back the indices of the transmitters that should participate in the transmission. For each hop (of either single hop or multihop protocols), the transmitter transmits only when it receives feedback information containing its index number. Otherwise, the node is silent. Moreover, transmitting nodes operate with fixed
transmitting power $P$ and fixed transmission rate $R_{0}$. In other
words, transmitters do not adapt their transmission power and rates
to instantaneous channel conditions. Suppose that at a particular
hop, a subset $\set{S}$ of the total nodes are scheduled for
transmission. Then, by definition, the transmission of node $i\in
\set{S}$ to receiving node $j$ is successful if the receiver SINR
(signal-to-interference-plus-noise ratio) exceeds a predefined
threshold $\beta_{0}$, i.e.,
\begin{equation}
\mathsf{SINR}_{i,j}=\frac{P\,\gamma _{i,j}}{\sigma_{N}^{2}+P\sum_{\substack{
t\neq i \\ t\in \set{S}}}\gamma_{t,j}}\geq \beta _{0},
\label{eq:general_SINR}
\end{equation}%
where $\sigma_{N}^{2}$ denotes the power of noise which is assumed
to be circular symmetric Gaussian, and $P$ is the transmission power
common to all transmitting nodes. For simplicity, we assume that the
fixed transmission rate $R_{0}$ and the threshold $\beta _{0}$ are
related by Shannon's equation $R_{0}=\log (1+\beta _{0})$.

In Section~\ref{sec:proof_upper_bound}, throughput upper bounds, which hold for any realization of the fading processes, are derived.\footnote{Note that the throughput upper bounds (cf. Theorem~\ref{thm:upper_bound}) derived in Section~\ref{sec:proof_upper_bound}, hold for \emph{any realization} of the fading processes, and thus are stronger than the \emph{average throughput} bounds of Sections~\ref{sec:achieve_square_root_n} and ~\ref{sec:beyond_squre_root_n}.} In Section~\ref{sec:achieve_square_root_n} and Section~\ref{sec:beyond_squre_root_n},
achievable average throughputs for a specific scheme are analyzed under specific fading distributions. As shown in ensuing sections, under the aforementioned assumptions concerning the transmitting and receiving strategies, the system throughput is fundamentally limited by interference.

\section{Upper Bound of Throughput Scaling for Distributions with Finite
Mean and Variance}

\label{sec:proof_upper_bound}

\setcounter{theorem}{0} 

In this section, we derive throughput scaling upper bounds for
various communication protocols under the assumption that the
channel connection distribution has \emph{finite mean} and
\emph{finite variance}. This class of distributions encompasses a
rich range of wireless channel models, e.g., Rayleigh, lognormal,
and Nakagami, etc., as special cases.

\begin{theorem}
\label{thm:upper_bound} Under the channel and system models
formulated in Section~\ref{sec:sysmod}, and for a class of channel
connection distributions with finite mean and variance, the
throughput scaling is upper-bounded by $O(n^{1/3})$ for single-hop
schemes, and $O(n^{1/2})$ for two-hop schemes. Furthermore, multihop
schemes cannot achieve better scaling than two-hop schemes.
\end{theorem}

The proof relies on a genie-aided scheme which is assumed to have
the capability of performing an exhaustive search over all
transmitting scenarios and therefore can yield the maximum number of
concurrent successful
transmissions.\footnote{%
Recalling that since transmission rate is fixed, maximizing the
number of concurrent, successful transmissions is tantamount to
maximizing the system throughput.} Consequently, throughput upper
bound is established by showing that, with the probability
approaching one, the genie-aided scheme cannot achieve an
arbitrarily large number of concurrent successful transmissions
between S--D pairs.

\begin{IEEEproof}[Proof of Theorem \ref{thm:upper_bound}]
We present the detailed proof of the throughput upper bound for the
single-hop scheme. Proofs for the two-hop and multihop schemes
follow similarly.

\emph{Throughput Upper Bound of Single-hop Schemes:} As illustrated
in Fig.~\ref{fig:single_hop_ex}, in the single-hop scheme, source
nodes communicate directly with destination nodes.
\begin{figure}[h]
\centering
\includegraphics{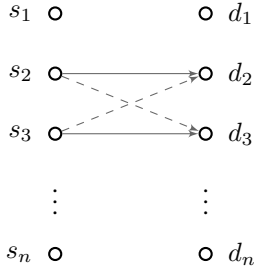}
\caption{Illustration of selecting a $2$-element active S--D set.
Here, only the second and third S--D pairs are active, while the
others are silent.} \label{fig:single_hop_ex}
\end{figure}

Consider the following genie-aided scheme, which performs an
exhaustive search over all $\binom{n}{m}$ possible ways of grouping
$m$ S--D pairs, and labels the set as \emph{valid} if all $m$ S--D
transmissions in the set are successful. Let $X(m)$ be the number of
valid sets. For certain value of $m$, $X(m)\geq 1$ implies that
$mR_0$ total throughput is achievable. On the other hand, $\Pr
[X(m)\geq 1] \to 0$ implies that $mR_0$ is not supportable in the
probabilistic sense. (Note that the proof method enforces the condition that $\sum_{m'>m}^{n}\Pr[X(m')\geq 1]$ decreases to zero exponentially with $n$; see \cite{AS:00} for a systematic treatment of this probabilistic method of proof.) To proceed, write $X(m)$ as a sum of indicator
functions:
\begin{align}
X(m)& =\sum_{\substack{ \set{S}\subset \{1,\ldots ,n\}  \\ \abs{\set{S}}=m}}%
\indicator\Bigl(\sinr_{i}\geq \beta _{0},\;\forall i\in \set{S}\Bigr) \\
& =\sum_{\substack{ \set{S}\subset \{1,\ldots ,n\}  \\ \abs{\set{S}}=m}}%
\indicator\Biggl(\frac{\gamma _{i,i}}{\frac{1}{\rho }+\sum_{\substack{ j\neq
i  \\ j\in \set{S}}}\gamma _{j,i}}\geq \beta _{0},\;\forall i\in \set{S}%
\Biggr),
\end{align}%
where $\beta_{0}$ is, as before, the SINR threshold, $\rho
\triangleq P/\sigma_{N}^{2}$ denotes the average SNR of the S--D
link. Then, using Markov's inequality, $\Prob{X(m)\geq 1}$ can be
upper-bounded as follows:
\begin{eqnarray}
\lefteqn{\Prob{X(m)\geq 1}} \notag \\
&& \leq \expt{X(m)}  \label{eq:markov_inq} \\
&& =\binom{n}{m}\Biggl(\PP\biggl[\frac{\gamma_{i,i}}{\frac{1}{\rho
}+\sum
_{\substack{ j\neq i  \\ j\in \set{S}}}\gamma_{j,i}}\geq \beta _{0}\biggr]%
\Biggr)^{m},  \label{eq:first_moment_ineq}
\end{eqnarray}%
where \eqref{eq:first_moment_ineq} follows from the linearity of the
expectation, and the fact that in our i.i.d. channel model,
different $\sinr_{i}$s involve independent channel gains, and thus
the terms $\sinr_{i}$s are i.i.d. over $i\in \set{S}$.

The approach to find the largest $m$ for which communication is
successful is to find an expression of the upper bound that is a
function of $m,$ and subsequently study its behavior as a function
of $m$. As detailed in the
Appendix~\ref{app:derive_Pr_Sm}, as far as scaling law is concerned,
$\Pr\Bigl[\frac{\gamma_{i,i}}{\frac{1}{\rho}+\sum
_{j\neq i,\,j\in \set{S}}\gamma_{j,i}}\geq \beta _{0}\Bigr]$ is upper-bounded by $O(1/m^2)$. In the sequel,
without loss of optimality, we write $\Pr\Bigl[\frac{\gamma_{i,i}}{\frac{1}{\rho}+\sum
_{j\neq i,\,j\in \set{S}}\gamma_{j,i}}\geq \beta _{0}\Bigr]\leq c_1/m^{2}$ for some constant $c_1$. Then, it follows from
\eqref{eq:first_moment_ineq} that
\begin{align}
\Prob{X(m)\geq 1}& \leq \binom{n}{m}\Bigl(\frac{c_1}{m^{2}}\Bigr)^{m}  \notag
\\
& \leq \Bigl(\frac{n\cdot e}{m}\Bigr)^{m}\Bigl(\frac{c_1}{m^{2}}\Bigr)^{m}
\label{eq:single_hop_temp1} \\
& =e^{m(\log (c_1e\cdot n)-3\log m)},  \label{eq:first_moment_ineq2}
\end{align}%
where \eqref{eq:single_hop_temp1} follows from the inequality
$\binom{n}{k}\leq (\frac{ne}{k})^{k}$.

On setting $m=(1+\varepsilon )(c_1e)^{1/3}\,\,n^{1/3},$ we have for
any $\varepsilon >0$,
\begin{align}
\label{eq:Pr_decreses_Exp}
\Prob{X(m)\geq 1}& \leq e^{-3(1+\varepsilon )\log (1+\varepsilon
)(c_1e)^{1/3}\,n^{1/3}} \\
& \rightarrow 0\qquad \text{as $n\rightarrow \infty $},
\end{align}%
which implies that, with probability approaching $1$, we cannot find
a valid set with $m=(1+\varepsilon )(c_1e)^{1/3}\,\cdot \,n^{1/3}$
S--D pairs. In the terminology of throughput scaling laws, the
system throughput is upper-bounded by $O(n^{1/3})$. \smallskip

\emph{Throughput Upper Bound of Two-hop Schemes:} Next, we consider two-hop
schemes, in which communication from sources to their destinations takes
place in two-hop fashion, with the help of $m$ relay nodes, as illustrated
in Fig.~\ref{fig:two_hop_ex}.

\begin{figure}[h]
\centering
\includegraphics{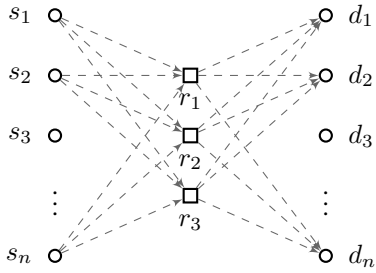}
\caption{Illustration of a two-hop scheme with $m=3$ relay nodes.
Sources $\{1,2,n\}$ form a $3$-element active source set in the
first hop. Destinations $\{1,2,n\}$ form a $3$-element active
destination set in the second hop.} \label{fig:two_hop_ex}
\end{figure}

In the following discussion, we assume that the fixed transmission
rate is greater than or equal to $1$ bit/s/Hz (i.e., $R_{0}\geq 1$),
and thus SINR $\geq 1 $ is required at the receiver for successful
communication. This additional assumption rules out the possibility
that multiple transmitters
can be successfully decoded by the same receiver.\footnote{%
This assumption allows us to simplify the analysis. We expect our
results to hold also for $R_{0}<1$.} Accordingly, if $m$ relay nodes
are deployed, the maximum number of concurrent successful
transmissions is at most $m$.

We start with the first hop.\footnote{By considering the first and
second hop separately, we implicitly assume that relay nodes are
capable of buffering data from all source nodes such that the relays
always have packets for transmission when being scheduled in the
second hop \cite{GT:02}.} Two-hop schemes differ from single-hop schemes in that
they have more freedom in associating transmitter-receiver pairs.
For example, in a two-hop scheme, each source can be decoded by any
relay, since any relay can subsequently send the message to the
intended destination. This is in contrast to single-hop schemes, in
which a source must be decoded by the destined destination node.
When $m$ sources are scheduled to transmit to $m$ relays, there are
$m!$ ways of associating the source-to-relay (S--R) pairs. Thus, if
we define $X(m)$ as the number of valid sets of $m$ concurrent S--R
transmissions, we have (cf.~\eqref{eq:first_moment_ineq})
\begin{equation*}
\Prob{X(m)\geq 1}\leq \binom{n}{m}\,m!\,\bigl(\PP [ \sinr \geq
\beta_0]\bigr)^{m}.
\end{equation*}%
In words, to test whether $m$ concurrent S--R transmissions are
possible (or equivalently if a throughput of $mR_{0}$ is
achievable), a genie can search over all $\binom{n}{m}$ ways of
selecting $m$-element sets of sources. Moreover, for each selection
of an $m$-element set, one has $m!$ ways of associating the sources
to the relays. Then, we have a total of $\binom{n}{m}\,m!$ ways of
determining S--R pairs. Similar to the single-hop case, the
probability of all $m$ transmissions being successful is given by
$(\Pr [\sinr\geq \beta _{0}])^{m}$.

Analogous to \eqref{eq:single_hop_temp1} and \eqref{eq:first_moment_ineq2}, we
have
\begin{align*}
\Prob{X(m)\geq 1}& \leq \binom{n}{m}m!\Bigl(\frac{c_1}{m^{2}}\Bigr)^{m} \\
& =\frac{n!}{(n-m)!}\Bigl(\frac{c_1}{m^{2}}\Bigr)^{m} \\
& \leq \Bigl(c_1\frac{n}{m^{2}}\Bigr)^{m} \\
& =e^{m(\log n-2\log m+\log c_1)}.
\end{align*}%
On setting $m=(1+\varepsilon )c_1^{1/2}n^{1/2}$, it is straightforward to show
that $\Prob{X(m)\geq 1}\rightarrow 0$. This implies that, with probability
approaching $1$, one cannot have $(1+\varepsilon )c_1^{1/2}n^{1/2}$ concurrent
successful transmissions. Equivalently, the system throughput is
upper-bounded by $(1+\varepsilon )c_1^{1/2}n^{1/2}$. In terms of scaling laws,
the throughput is $O(n^{1/2})$. Compared to the single-hop setup, we see
that the additional freedom of associating transmitter--receiver pairs leads
to a higher upper bound on the throughput.

According to the scheduling policy of the genie scheme, the second
hop is similar to the first hop. Specifically, in the second hop, we
seek among all $\binom{n}{m}\,m!$ possible ways of relaying the
information from the relays to the destinations (R--D), the mappings
such that all involved transmissions are successful. This is
mathematically equivalent to the first hop, and therefore it results
in the same upper bound on the throughput.

Combining the first hop and the second hop, we conclude that the throughput
of two-hop schemes as a whole is $O(n^{1/2})$. \smallskip

\emph{Throughput Upper Bound of Multihop Schemes:} In multihop schemes, the
communication between sources and destinations is facilitated by multiple
layers of relaying.

We have discussed previously that a single layer of relays is
beneficial in that it offers flexibility in routing messages between
sources and destinations. However, in the random connection model
stipulated in this work, there is no advantage in having multiple
layers of relays. This is because in a multihop scheme, the
throughput is limited by the layer with minimal throughput. Adding
more relay layers does not resolve this bottleneck, and it offers no
additional flexibility compared to the two-hop scheme. Therefore, as
far as throughput is concerned, multihop schemes cannot do better
than two-hop schemes.

This completes the proof of Theorem \ref{thm:upper_bound}.\newline
\end{IEEEproof}

Several remarks are in order.

\begin{remark}
The theorem establishes a universal upper bound on the throughput
that is valid for a class of fading distributions with finite mean
and variance. This class encompasses a number of familiar models in
wireless communications, e.g., Rayleigh, lognormal, and Nakagami,
among others. Perhaps surprisingly, particular channel model with
finite mean and variance may yield the throughput scaling behavior
drastically different than the upper bound of
Theorem~\ref{thm:upper_bound}. For instance, it is shown in
\cite{CHSP:07} that for a Rayleigh fading environment, the
throughput scales as $\Theta (\log n)$, which is clearly recognized
to grow much slower than $O(n^{1/2})$. Thus, in general, the
throughput scaling upper bound for specific distributions should be
studied on a case-by-case basis. Fortunately, for achievable throughput, insights of the throughput scaling behavior can be gained. See Remark~\ref{rmk:multiuser_diversty_interpretation} for discussion.
\end{remark}

\begin{remark}
It is important to note that the conclusion that multihop schemes
are not beneficial stems directly from the dense network and random
fading model adopted in our work. (A similar conclusion is drawn
also in \cite{GHH:06}.) In other different setups, however, multihop schemes might
be beneficial. For example, in an extended network with path-loss attenuation,
multiple hopping is necessary due to the limited transmission range
of each hop. As a matter of fact, the nearest neighbor multihop is
order-optimal in achieving the best throughput scaling in the
geometrical path-loss attenuation model with attenuation factor
$\alpha \geq 3$ \cite{OLT:07}.
\end{remark}

\section{Achievability of $\Theta (n^{1/2})$ Throughput}
\label{sec:achieve_square_root_n}

In the preceding section, we established the throughput upper bounds
for different communication protocols by finding such $m$ that $\Pr [X(m)\geq 1]\rightarrow 0$. We have proved that, in the
random connection channel model, two-hop schemes have a higher
throughput upper bound than single-hop schemes. However, it is not
clear from the preceding discussion whether this upper bound is
achievable. By achievable, we mean that the scheme can operate with
distributed scheduling, i.e., without the genie's clairvoyance. In this section, we establish the achievability result of $\Theta(n^{1/2})$ for two-hop relaying schemes by a constructive example.

We formulate the achievability result in the following theorem:

\smallskip

\begin{theorem}
\label{thm:opportunistic_by_MD} Under the same assumptions as
Theorem~\ref{thm:upper_bound}, an average throughput scaling of $\Theta
(n^{1/2})$ is achievable by a two-hop scheme.
\end{theorem}

\smallskip

This achievability result is established by the two-hop opportunistic
relaying scheme \cite{CHSP:07}. In \cite{CHSP:07}, the scheme has been shown to
achieve $\Theta(\log n)$ throughput scaling under the Rayleigh fading model. In this section, we will see that the opportunistic relaying scheme is able to achieve
$\Theta(n^{1/2})$ throughput scaling under the specific fading model of \eqref{eq:mysterious_fading}. Before proceeding to present the
proof of Theorem~\ref{thm:opportunistic_by_MD}, we need first to set the
stage by briefly reviewing the two-hop opportunistic relaying scheme and by outlining the generic throughput expressions valid for any fading distributions (whereas the derivations in \cite{CHSP:07} are tailored to the Rayleigh fading model).
Subsequently, we introduce the channel model used to prove the achievability
result, and finally complete the proof.

\subsection{Two-Hop Opportunistic Relaying Scheme}

\label{sec:two_hop_opp_schme}

The two-hop opportunistic relaying scheme was first proposed in
\cite{CHSP:07}. The key idea is to exploit the multiuser diversity
of the network and apply it to compensate for the effect of
concurrent interference sources. To enable the channel-dependent
scheduling, low rate feedback is required from receivers to
transmitters. For the sake of completeness, the scheduling policy of
the scheme and the corresponding throughput expressions are briefly
reviewed in the sequel. See \cite{CHSP:07} for details.

\subsubsection{System Model}

\begin{figure*}[tbh]
\centerline{ \subfigure[]{\includegraphics{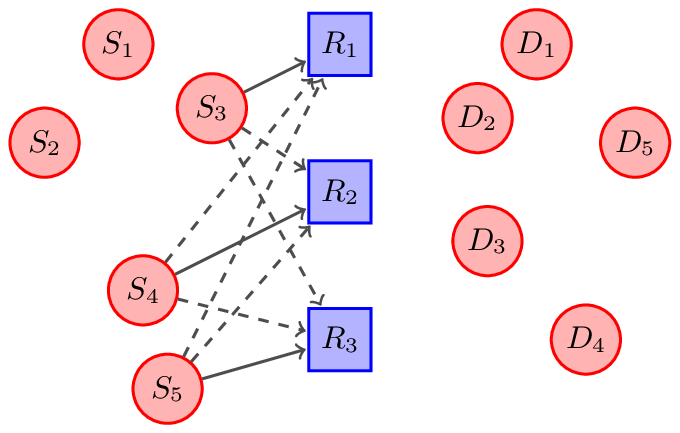}}
\hfil \subfigure[]{\includegraphics{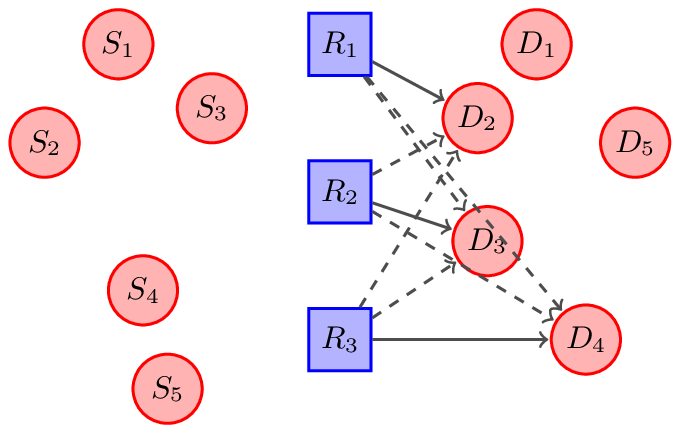}}}
\caption{A two-hop network with $n=5$ S--D pairs and $m=3$ relay
nodes (denoted by the blue squares). (a) In the first hop, source
nodes $\{3,4,5\}$ transmit to the relays. (b) In the second hop, the
relays transmit to the destination nodes $\{2,3,4\}$. Solid lines
indicate scheduled links, while dashed lines indicate interference.}
\label{fig:sys_model}
\end{figure*}
As illustrated in Fig.~\ref{fig:sys_model}, we consider a wireless
network with $n$ S--D pairs and $m$ relay nodes operating in a
two-hop decode-and-forward fashion. Since we are interested in an
achievable result, without loss of generality, we further assume a
fixed transmission rate $R_{0}=1$ bit/s/Hz.

\subsubsection{Scheduling and Throughput}

\label{sec:scheduling}

\emph{\newline \indent First Hop Scheduling and Throughput:} The
first hop scheduling can be thought of as a natural generalization
of the classic multiuser diversity scheme with single receiver
antenna \cite{KH:95} to multiple, decentralized antennas.
Specifically, denote the channel connection from $i$th source
$(1\leq i\leq n)$ to $j$th relay $(1\leq j\leq m)$ as $\gamma
_{i,j}$, then relay $j$ schedules its best source $i^{\ast }=\arg
\max_{i}\gamma _{i,j},$ by feeding back the index $i^{\ast }$. Since
there are $m$ relays, and each of them schedules one source for
transmission, so up to $m$ sources can be scheduled at the same
time. By accounting only for the scenario when $m$ distinct sources
are scheduled (denoting such event as $N_{m}$ with
$\Pr[N_m]=n(n-1)\cdots (n-m+1)/n^{m}$ \cite{CHSP:07}), a lower bound
on throughput of the first hop $R_{1}$ can be written as,
\begin{align}
R_{1}& \geq \Pr [N_{m}]\sum_{j=1}^{m}\Pr [\mathsf{SINR}_{i^{\ast
}\!\!,j}\geq 1]\cdot 1 \label{eq:rhs_r1}\\
&= c_2 \sum_{j=1}^{m}\Pr [\mathsf{SINR}_{i^{\ast }\!\!,j}\geq
1] \quad \text{ for some $0<c_2\leq 1$}
\label{eq:Pr_Nm_to_1} \\
&=c_2 \,m\cdot \Pr \biggl[\frac{\gamma _{i^{\ast}\!\!,j} }{\frac{1}{\rho
}+\sum
_{\substack{ i\neq i^{\ast}  \\ i\in \mathcal{S}}}\gamma _{i,j}}\geq 1 %
\biggr],  \label{eq:lb_R1}
\end{align}%
where where right-hand side (RHS) of \eqref{eq:rhs_r1} is due to the
fact that, conditioned on $N_m$, there are total $m$ concurrent
transmissions, each transmitting $1$ bit/s/Hz with a probability of success of $\Pr[\mathsf{SINR}\geq 1]$; \eqref{eq:Pr_Nm_to_1}
follows from the fact that $\Pr [N_{m}]= \Theta(1)$ with
$m=O(n^{1/2})$; and
\eqref{eq:lb_R1} follows from the fact that $\mathsf{SINR}_{i^{\ast
}\!\!,j}$s for different $j$'s involve i.i.d. channel realizations,
and thus are i.i.d. To see $\Pr[N_m]=\Theta(1)$ as $m=O(n^{1/2})$, we can show that the lower bound on $\Pr[N_m]$ is $\Theta(1)$ as follows:
\begin{align}
\Pr[N_m]&=\frac{n}{n}\frac{n-1}{n}\cdots\frac{n-m+1}{n} \notag \\
&\geq \left(\frac{n-m+1}{n}\right)^m \notag \\
&\geq \left(1-\frac{c_3\sqrt{n}-1}{n}\right)^{c_3\sqrt{n}}  \notag
\intertext{with $m=c_3\sqrt{n}$ for some constant $c_3$}
&\to e^{-c_3^2} \label{eq:e_xn}\\
&=\Theta(1), \notag
\end{align}
where \eqref{eq:e_xn} follows from the fact that $(1-x)^n\to e^{-nx}$ for $x\to 0$ and $n\to \infty$. Since $\Pr[N_m]\leq 1$, without loss of generality, we denote  $\Pr[N_m]$ by a constant $0<c_2\leq 1$ in \eqref{eq:Pr_Nm_to_1}.

\emph{Second Hop Scheduling and Throughput:} In the second hop, $m$
relay nodes communicate with destination nodes. The scheduling of
the second hop works as follows. Each destination node $j$ computes
$m$ SINRs (cf.~\eqref{eq:general_SINR}) by assuming that relay $k$
($1\leq k\leq m$) is the desired sender, and all other $(m-1)$
relays generate interference. If the destination node $j$ captures
one good SINR, say $\mathsf{SINR}_{k^{\ast},j}\geq 1$ for some
$k^{\ast}$, it instructs relay $k^{\ast}$ to send data by feeding
back the relay index $k^{\ast}$. Otherwise, the node $j$ does not
provide feedback. Since the scheduling of the second hop depends on
$\mathsf{SINR}\geq 1,$ by definition, all transmissions at $1$
bit/s/Hz will be successful. Consequently, the average throughput of
the second hop depends on whether relay nodes are scheduled (by a
destination node that measured SINR $\geq 1)$. Given the scheduling
criteria as above, we have the throughput of the second hop $R_{2}$
expressed as follows
\begin{align}
R_{2}& =\sum_{k=1}^{m}\Pr [\text{relay $k$ is scheduled}]
\notag\\
& =m\Pr [\text{relay $k$ is scheduled}]  \label{eq:R2_step1} \\
& =m\Pr \bigl[\cup_{j=1}^{n}\text{relay $k$ receives feedback from
$j$}\bigr] \notag
\\
& \geq m\Pr [\text{relay $k$ receives feedback from $j^{\ast }$}]
\label{eq:lb_R2_step1} \\
&=m\cdot \Pr \biggl[\frac{\gamma _{k,j^{\ast}} }{\frac{1}{\rho
}+\sum _{\substack{ k^{\prime}\neq k  \\ 1\leq k^{\prime}\leq
m}}\gamma _{k^{\prime}\!,j^{\ast}}}\geq 1 \biggr],  \label{eq:lb_R2}
\end{align}%
where, again, \eqref{eq:R2_step1} follows from our i.i.d. model and
the lower bound of \eqref{eq:lb_R2_step1} follows from the fact we
account for the event only when the best node, i.e., $j^{\ast
}\triangleq \arg \max_{j}\gamma_{k,j}$, requests transmission.

\begin{remark}
It is noted that the scheduled policies for both the first hop and
the second hop are of opportunistic nature. Consequently, it is
possible that the active destination nodes in a time-slot (at the
second hop) are not exactly \emph{those} that correspond to the
active source nodes in a previous time-slot (as deliberately
exemplified in Fig.~\ref{fig:sys_model}). Hence, the relays are
assumed to have the ability to buffer the data received from source
nodes, such that it is available when the opportunity arises to
transmit it to the intended destination nodes over the second hop of
the protocol \cite{GT:02}. This ensures that relays always have packets destined
to the nodes that are scheduled. With the help of this assumption,
the throughput of the two-hop scheme is defined as $R\triangleq
\frac{1}{2}\min \{R_{1},R_{2}\},$ where $1/2$ accounts for the fact
that transmitters transmit only half of time.
\end{remark}

\begin{remark}
\label{rmk:multiuser_diversty_interpretation}
It is enlightening to look at the role played by the channel model in the throughput expressions \eqref{eq:lb_R1} and
\eqref{eq:lb_R2}. The specific channel model determines the desired
signal power (the numerator) as well as the interference power.
However, we know from the law of large numbers that the interference
term approaches its mean value, which scales linearly
with the number of relays $m$. However, the signal
power is more sensitive to channel models (i.e., different models
have different multiuser diversity gains). For example, the multiuser
diversity gain scales as $\log n$ for Rayleigh fading channels
\cite{VTL:02}, but scales as $e^{\sqrt{2\log n}\,\sigma}$ for
lognormal fading channels \cite[Sec.~2.3.3., p.~67]{Galambos:78}. In order to have non-diminishing data rate, we need accordingly non-diminishing SINR. Therefore, the limiting operating scenario would be that the desired signal strength (i.e., the multiuser diversity gain) is of the same order as the aggregate interference power. This implies that \textit{the maximal number of scheduled transmissions (and hence the maximal throughput) is governed by the order of multiuser diversity gain}. This intuition is explored numerically in \cite{CH:08} for various fading models.
\end{remark}

\subsection{Channel Model}

\label{sec:channel_model}

Given the intuition from Remark~\ref{rmk:multiuser_diversty_interpretation},
for $\Theta(n^{1/2})$ throughput scaling, a fading distribution is required that has a multiuser diversity gain of the order of $\Theta(n^{1/2})$. To this end, we assume the channel (power) gains (S--R and R--D) are drawn
independently from
the following cdf:%
\begin{gather}
F_{n}(x)=\biggl(\frac{(x-\mu )/\sigma }{\sqrt{2n-1}}\frac{n-1}{n}+%
\frac{1}{n}\biggr)^{1/(n-1)},  \label{eq:CDF} \\
\mu -\tfrac{\sqrt{2n-1}}{n-1}\sigma \leq x\leq \mu +\sqrt{2n-1}\sigma ,
\notag
\end{gather}%
where $\mu $ and $\sigma $ are respectively, the mean and variance
of the fading connection, and are assumed to be finite.

Indeed, the cdf in \eqref{eq:CDF} has the
unique property that it maximizes the expectation of the extreme
value among all distributions possessing finite mean and variance
\cite{Gumbel:54}. Moreover, the mean is given by
$\mathbb{E}[M_{n}]=\mu +\sigma \frac{n-1}{\sqrt{2n-1}}$, where
$M_{n}$ denotes the maximum of $n$ i.i.d. random variables having the
common distribution of \eqref{eq:CDF}.

In what follows, we assume $\mu =\sigma =1$. This simplification
incurs no loss of optimality with respect to the scaling behavior.
The following property, easily verified (see
Appendix~\ref{app:Pr_Mn_geq_EMn} for derivation), will be useful in
the ensuing discussion:
\begin{equation}
\label{eq:Pr_Mn_geq_EMn}
\Pr \left[ M_{n}>\frac{n-1}{\sqrt{2n-1}}\right] =\Pr \left[ M_{n}\leq \frac{%
n-1}{\sqrt{2n-1}}\right] =\frac{1}{2}.
\end{equation}

\begin{figure*}[!b]
\vspace*{4pt}
\hrulefill
\normalsize
\setcounter{MYtempeqncnt}{\value{equation}}
\setcounter{equation}{22}
\begin{align}
\Pr & [\mathsf{SINR}\geq 1]\approx \PP \biggl[\frac{M_{n}}{I_{m-1}}\geq 1\biggr]\label{eq:temp1} \\
& =\underbrace{\PP\biggl[\frac{M_{n}}{I_{m-1}}\geq 1\bigg\vert %
M_{n}>m,I_{m-1}\leq m\biggr]}_{=1}\Pr [M_{n}>m,I_{m-1}\leq m]  \notag \\
& \quad +\underbrace{\PP\biggl[\frac{M_{n}}{I_{m-1}}\geq 1\bigg\vert %
M_{n}>m,I_{m-1}>m\biggr]}_{\geq 0}\Pr [M_{n}>m,I_{m-1}>m]  \notag \\
& \quad +\underbrace{\PP\biggl[\frac{M_{n}}{I_{m-1}}\geq 1\bigg\vert
M_{n}\leq
m,I_{m-1}\leq m\biggr]}_{\geq 0}\Pr [M_{n}\leq m,I_{m-1}\leq m]  \notag \\
& \quad +\underbrace{\PP\biggl[\frac{M_{n}}{I_{m-1}}\geq 1\bigg\vert
M_{n}\leq
m,I_{m-1}>m\biggr]}_{=0}\Pr [M_{n}\leq m, I_{m-1}>m]  \notag \\
& \geq \Pr [M_{n}>m,I_{m-1}\leq m]  \label{eq:temp2} \\
& \geq \frac{1}{26},  \label{eq:temp3}
\end{align}%
\setcounter{equation}{\value{MYtempeqncnt}}
\end{figure*}

\subsection{Proof of Achievability}
\label{sec:achieve_square_n}

Before embarking on the achievability proof, we first characterize the
aggregate interference stemming from all $(m-1)$ concurrent
transmitting sources, $\sum_{i\neq i^{\ast}, \,i\in \mathcal{S}}\gamma _{i,j}$, in the expressions of \eqref{eq:lb_R1} and
\eqref{eq:lb_R2}. It is shown in \cite{CHSP:07} that, in the regime of large $n$, the $\{\gamma_{i,j}\}$ are asymptotically independent. Denote $I_{m-1}\triangleq \sum_{i\neq i^{\ast}, \,i\in \mathcal{S}}\gamma _{i,j}$. In the ensuing
achievability proof, we need to show that $\Pr[I_{m-1}\leq m]$ is at least greater than zero. To show this, we need the following technical lemma due to Feige \cite[Th.~1]{Feige:06}.

\begin{lemma}
Let $X_1,\ldots,X_N$ be arbitrary nonnegative independent random variables, with expectations $\expt{X_i}\leq 1, \,\forall \,i$. Let $S=\sum_{i=1}^{N}X_i$. Then for every $\delta>0$,
\begin{equation}
\label{eq:Feige}
\Pr\bigl[S \leq \expt{S}+\delta \bigr] \geq \min\left(\frac{\delta}{1+\delta},\frac{1}{13}\right).
\end{equation}
\end{lemma}

In our case, from the linearity of expectation, we have $\expt{I_{m-1}}=m-1$. Thus, from \eqref{eq:Feige} we have:\footnote{Note that the factor $\frac{1}{13}$ in \eqref{eq:Feige} may be too conservative and Feige conjectured in \cite{Feige:06} that $\frac{1}{13}$ can be replaced by $\frac{1}{e}$.}
\begin{equation}
\label{eq:Pr_I_m_minus_1}
\Pr\bigl[I_{m-1} \leq m \bigr] \geq \min\left(\frac{1}{2},\frac{1}{13}\right)=\frac{1}{13}.
\end{equation}

The proof of Theorem~\ref{thm:opportunistic_by_MD} follows now.
\smallskip

\begin{IEEEproof}[Proof of Theorem \ref{thm:opportunistic_by_MD}]
Both lower bounds \eqref{eq:lb_R1} and \eqref{eq:lb_R2} manifest a
tradeoff between data rate and interference: Increasing the number
of relay nodes is beneficial from the rate's perspective, but the
interference term is also increased. Since the operating SINR
threshold is one, a good operating point for $m$ should be such that
the SINR is of the order of one. Hence, noting the fact that the
mean value of the signal power is given by $\mathbb{E}[M_{n}]=1+\frac{n-1}{\sqrt{%
2n-1}}$, we set the number of relay nodes $m$ to
$\frac{n-1}{\sqrt{2n-1}}$. This specific setup causes $\Pr
[\mathsf{SINR}\geq 1]$ in \eqref{eq:lb_R1} and \eqref{eq:lb_R2} to
be of the order of one. Specifically, in the derivation, which is shown at
the bottom of the page, the approximation in \eqref{eq:temp1} is due to the fact that
Gaussian noise is ignored (note that the system is interference
limited); the inequality in \eqref{eq:temp2} follows since one
combination of $M_{n}$ and $I_{m-1} $ is accounted for in applying
the total probability theorem; and \eqref{eq:temp3} follows from the
fact that $M_{n}$ and $I_{m-1}$ are asymptotically independent
\cite{CHSP:07} and from \eqref{eq:Pr_I_m_minus_1}.
\addtocounter{equation}{3}

Together, $m=\frac{n-1}{\sqrt{2n-1}}$ and \eqref{eq:temp3} lead to
the conclusion that both \eqref{eq:lb_R1} and \eqref{eq:lb_R2} are
greater than or equal to $\frac{1}{26}\frac{n-1}{\sqrt{2n-1}}$, which
scales as $\Theta (n^{1/2})$ as $n\rightarrow \infty $. This
completes the proof of Theorem~\ref{thm:opportunistic_by_MD}.
\end{IEEEproof}
\smallskip

\begin{remark}
In this section, we have focused on the achievability of
$\Theta(n^{1/2})$ for two-hop schemes. To this end, we have
introduced the connection model of \eqref{eq:CDF}, which is a
function of $n$. While such a distribution is not likely to be encountered in practice,
the result of Theorem~2 is still of theoretical importance as it confirms
that the $O(n^{1/2})$ upper bound in Theorem~\ref{thm:upper_bound} is tight.
\end{remark}

Finally, it is important to note that the achievability of the result in this section is established for the specific two-hop opportunistic relaying scheme. In pursuing the achievable $O(n^{1/2})$ throughput, it is of both theoretical and practical interest to study other relaying schemes and examine their achievable throughputs under other channel models.

\section{Beyond $\Theta (n^{1/2})$}

\label{sec:beyond_squre_root_n}

The preceding sections restrict the discussion to a class of channel
connection distributions possessing finite mean and variance.
Theorems \ref{thm:upper_bound} and \ref{thm:opportunistic_by_MD}
state that, for such fading distributions, the throughput is
upper-bounded by $\Theta (n^{1/2})$. Now, a natural question to ask
is whether \emph{there exist channel models that can yield better,
even linear, throughput scaling?}

It turns out that the answer is affirmative. Again, since achievable
results are concerned, a case study suffices. To this end, we show
in this section that the two-hop opportunistic relaying scheme
(cf.~Section~\ref{sec:two_hop_opp_schme}) yields linear throughput
scaling under a class of fading models. This motivates in the second
part of the section, a broader discussion on channel models.

\subsection{Linear Throughput Scaling}

The key to attaining linear throughput scaling with the
opportunistic scheme is to achieve non-vanishing SINR
(cf.~\eqref{eq:general_SINR}) even when $n\rightarrow \infty $.
Intuitively, in order to achieve $\Theta(n)$ throughput scaling, one
has to schedule $\Theta(n)$ concurrent transmissions \emph{and} at
the same time to have $\Pr [\sinr \geq \beta_0]>0$ as $n\to \infty$.
We know from the discussion in
Section~\ref{sec:achieve_square_root_n} that the latter is not
possible if the underlying channel model has finite mean and
variance. This is because the interference increases linearly with
$n,$ but the signal power grows no faster than $n^{1/2}$.
Nevertheless, the following result shows that if we relax the
requirement of finite mean and variance, there exists a class of
fading models for which the signal power increases linearly with
$n$.

\begin{lemma}
\label{lem:Pr_Mn_In_geq_0} Let $X_{1},\ldots ,X_{n}$ be independent
random variables with common cdf $F$ satisfying
\begin{equation}
1-F(x)\sim \frac{c_0\cdot x^{-\nu }}{\Gamma (1-\nu )}
\label{eq:feller_distributions}
\end{equation}%
where $0<\nu <1$, $c_0$ is a constant, and $\Gamma (x)$ is the gamma
function. Define $I_{n}=X_{1}+\cdots +X_{n}$ and $M_{n}=\max
\{X_{1},\ldots ,X_{n}\}$; then
\begin{equation}
\Pr \biggl[\frac{M_{n}}{I_{n}}\geq 1-\nu\biggr]>0
\label{eq:good_SINR}
\end{equation}%
as $n\rightarrow \infty $.
\end{lemma}

\begin{proof}
See Appendix~\ref{app:prove_Pr_Mn_In_geq_0}.
\end{proof}

Note that \eqref{eq:good_SINR} holds independently of the system size. In other words, we can equivalently write $\Pr \bigl[\frac{M_{n}}{I_{n}}\geq 1-\nu\bigr]=c_4$ for some constant $c_4>0$ as $n\to \infty$. This is in contrast to the fading distributions that we have studied in Section~\ref{sec:proof_upper_bound}. There $\Pr[\mathsf{SINR}\geq \beta_0]$ decreases to zero as $m$ increases.

With the help of the lemma, one can set the system parameters of the
two-hop opportunistic relaying scheme as follows to achieve linear
throughput scaling. First, set SINR threshold $\beta_{0}=1-\nu$, and
the corresponding $R_{0}=\log (2-\nu)$. Then, set $m=n$ (i.e.,
deploy as many relay nodes as S--D pairs). Under this setup, it is
then straightforward to show that the total number of scheduled
transmissions is of the order of $n$, i.e.,
$\expt{\abs{\mathcal{S}}}=\Theta (n)$ (see
Appendix~\ref{app:show_Theta_n} for the derivation). This fact,
together with the fact that each of S--D pairs has a successful
transmission probability greater than zero
(cf.~\eqref{eq:good_SINR}), means that the system has linear
throughput scaling. We summarize this result in the following
theorem:

\smallskip

\begin{theorem}
\label{thm:linear_achiving_Pareto}A linear average throughput scaling of
$\Theta (n)$ is achievable by the two-hop relaying scheme under
fading models satisfying \eqref{eq:feller_distributions}.
\end{theorem}

\smallskip

\begin{example}
\label{ex:GT02_revisited} To show the application of
Theorem~\ref{thm:linear_achiving_Pareto}, it is illuminating to
revisit the famous Grossglauser--Tse scheme \cite{GT:02}, \emph{but}
under the framework of the random connection model.

In \cite{GT:02}, the authors considered a network model in which all
nodes are assumed to move in an i.i.d., stationary, and ergodic
fashion over a fixed area. The channel connections between nodes are
determined by a path-loss attenuation model, i.e., the connection
between any two nodes follows a path-loss law
$g(r)=\sqrt{G}\frac{1}{r^{\alpha}},$ where $r$ is the distance
between the nodes, $\alpha \geq 2$ is the path-loss exponent, and
$G$ is a constant depending on system parameters. It is shown that a
two-hop relaying scheme featuring nearest-neighbor communication can
achieve a linear throughput scaling. Note that the geometric
path-loss attenuation model used in \cite{GT:02} can be thought of
as a particular random fading model, where the randomness stems from
the random distance between nodes. For example, for each
transmitter, the received power seen by all receiving nodes are
random since distances are random. Following the approach of
\cite[Sec.~VIII.B]{GHH:06}, one can in fact characterize the
distribution of the received powers as follows. Consider a
simplified scenario where a source node at the center of a unit
area. All receiver nodes are assumed to be uniformly distributed
over the area. The node has transmit power $P$ and we denote $\gamma
$ the corresponding received power. Now, for a typical receiver, the
probability of receiving power less than $\gamma P$ is given by
\begin{equation}
F(\gamma )=1-\pi \gamma ^{-\frac{2}{\alpha}}  \label{eq:decay_cdf}
\end{equation}%
where $\gamma \in \lbrack \pi ^{\frac{\alpha}{2}},\infty ]$. 
Now, we can interpret the original Grossglauser--Tse scheme \emph{as
though} it is in the framework of the random connection model with
fading distribution specified by \eqref{eq:decay_cdf}. Seen another way, if we allow mobility of nodes in a network with path-loss attenuation model, mobility
effectively blurs the distinction between near and far nodes and the network exhibits homogeneity. From the \emph{effective} channel connection's perspective, the network is equivalent to operating under random channel model \eqref{eq:decay_cdf}. Consequently, the linear throughput achievability of \cite{GT:02} can be
explained by noting the fact that the distribution of
\eqref{eq:decay_cdf} is of the form of
\eqref{eq:feller_distributions} (with $\nu=2/\alpha$ and
$c_0=\pi\,\Gamma(1-2/\nu)$). As a matter of fact, the distribution of
\eqref{eq:decay_cdf} is referred to as a Pareto distribution
\cite{Feller_vol2:71}. The Pareto distribution is known to have
infinite mean, when $2/\alpha \leq 1,$ and infinite variance when
$2/\alpha \leq 2$. This is consistent with
Theorem~\ref{thm:upper_bound}.
\end{example}

It is important to point out that, though linear throughput scaling
is achievable, it does not mean that the system performance is not
hindered by interference. The effect of interference is revealed in
that increasing the transmission power of each node does not lead to
an increase in throughput. What happens is that, thanks to the
unique property of the Pareto distribution, we can operate at the
boundary between
interference-limited and degrees of freedom-limited regimes.\footnote{%
In a system with $n$ degrees of freedom, the ratio of the throughput $R$ to
the logarithm of $\snr$ in the high SNR regime is given by $n$, i.e., $\lim_{%
\snr\rightarrow \infty }\frac{R}{\log \snr}=n$ \cite{ZT:03}.}

As a final remark, it should be noted that
Theorem~\ref{thm:linear_achiving_Pareto} provides just one type of
channel models that can yield linear throughput scaling, and it does
not preclude the existence of other distributions that can yield
linear throughput scaling.

\subsection{Discussion On The Channel Model}

\label{sec:model_discussion}

Unlike \cite{GK:00,GT:02,XK:04,OLT:07} (among many others) in which
geometric path-loss channel models are assumed, this paper adopts
the random connection model initially proposed in \cite{GHH:06}. In
the random connection model, there is no notion of location and
distance. Instead, the channel connections are modeled as random
variables. This model applies to scenarios where the network is
physically small such that path-loss is less relevant or where the
mechanism of automatic gain control (AGC) is used to mitigate
path-loss attenuation. This former scenario is partially justified
by the channel measurements in \cite{LCSV:07}. There, the authors
report that, in vehicle-to-vehicle communication environment, the
standard deviation of channel strength can be as large as $17.5$ dB.

On the other hand, when large-scale path-loss attenuation becomes
significant, the recent work \cite{OJTL:08} formulated the network
in terms of three system parameters: short-distance SNR (defined as
the SNR over the typical nearest neighbor distance), long-distance
SNR (defined as the SNR over the largest scale in the network), and
the power path-loss exponent. Depending on various combinations of
the values of these parameters, four fundamental operating regimes
of wireless networks have been identified. One interesting regime is
when the network is \emph{heterogeneous} in nature, namely, the
short-distance SNR is large but the long-distance SNR is small. It
turns out that optimal scheme in this regime follows the principle
of ``\emph{cooperate locally but multihop globally},'' that is, the
optimal scheme should cooperate locally within clusters of
intermediate size and then multihop across clusters to reach the
destination. Focusing on the communication within the cluster, we
note that the random connection model of this work is a natural
candidate.


\section{Concluding Remarks}

\label{sec:conclusion}

This work has focused on the throughput scaling laws of fading
networks in the limit of a large number of nodes. Motivated by considerations of CSI availability, we have considered the
case in which receivers have perfect CSI, but transmitters have only
partial CSI via receiver feedback. We have characterized the
architectural differences between various communication protocols
from the throughput standpoint. Specifically, it has been shown that
for a class of fading channels with finite mean and variance, the
upper bound on the throughput scaling is given by $O(n^{1/3})$ for
single-hop schemes, and by $O(n^{1/2})$ for two-hop (and multihop)
schemes. We have also shown that the two-hop opportunistic relaying
scheme achieves the $\Theta (n^{1/2})$ throughput scaling upper
bound. Furthermore, by relaxing the constraints of finite mean and
variance of the fading model, we have shown that linear throughput
scaling $\Theta (n)$ is achievable with Pareto-type fading models
with the two-hop opportunistic relaying scheme.

Some interesting issues remain open. First, are there other relaying schemes which achieve the $O(n^{1/2})$ throughput under more realistic channel models? Another important question that is not addressed in this paper is whether the $O(n^{1/3})$ throughput upper bound for a single-hop scheme is achievable or not. It is also interesting to characterize what happens if we alleviate the assumption of conventional single-user decoding at the
receivers (e.g., allowing interference cancelation and joint decoding techniques), but keep the CSI availability framework. Finally, while this work assumes dedicated relays (which have no intrinsic traffic demands), it is of interest to look at the
\emph{ad hoc} setup in which a dynamic subset of nodes (source and destination) act as relays.

\appendices

\section{Derivation of the Upper Bound of $\Pr\Bigl[\frac{\gamma_{i,i}}{\frac{1}{\rho}+\sum
_{j\neq i,\,j\in \set{S}}\gamma_{j,i}}\geq \beta _{0}\Bigr]$}
\label{app:derive_Pr_Sm}

In this appendix, the upper bound on $\Pr\Bigl[\frac{\gamma_{i,i}}{\frac{1}{\rho}+\sum
_{j\neq i,\,j\in \set{S}}\gamma_{j,i}}\geq \beta _{0}\Bigr]$ is derived in the asymptotic regime of large $m$. For notational
convenience, let $Z\triangleq \sum_{\substack{ j\neq i  \\ j\in \set{S}}}%
\gamma _{j,i}$. Starting with the law of total probability, for any $s>0$, we have
\begin{eqnarray}
\lefteqn{\Pr \left[ \frac{\gamma_{i,i}}{1/\rho +Z}\geq \beta _{0}\right]} \notag\\
&&= \Pr \left[ \frac{\gamma _{i,i}}{1/\rho +Z}\geq \beta _{0} \biggl| Z\leq s\right]\Pr[Z\leq s] \notag \\
&&\quad +\Pr \left[ \frac{\gamma _{i,i}}{1/\rho +Z}\geq \beta _{0} \biggl| Z> s\right]\Pr[Z> s] \label{eq:total_Pr}\\
&&\leq \Pr[Z\leq s] + \Pr \left[ \frac{\gamma _{i,i}}{1/\rho +Z}\geq \beta _{0} \biggl| Z> s\right] \label{eq:two_Pr_leq_one}\\
&&\leq \Pr[Z\leq s] + \Pr \left[ \frac{\gamma _{i,i}}{1/\rho +s}\geq \beta _{0}\right], \label{eq:each_one_tbb}
\end{eqnarray}
where \eqref{eq:two_Pr_leq_one} follows from the fact that $\Pr \Bigl[ \frac{\gamma _{i,i}}{1/\rho +Z}\geq \beta _{0} \bigl| Z\leq s\Bigr]\leq 1$ and
$\Pr[Z> s]\leq 1$. Now, we proceed to upper-bound each term in \eqref{eq:each_one_tbb}.

To bound $\Pr[Z\leq s]$, we need the following large deviation result due to Maurer \cite[Th.~2.1]{Maurer:03}.

\begin{theorem}
Let $X_1,\ldots,X_N$ be independent real-valued random variables with $X_i\geq 0$ and $\expt{X_i^2}<\infty$ for $i=1,\ldots,N$. Set $S=\sum_{i=1}^{N}X_i$ and let $x>0$. Then
\begin{equation}
\Pr[S\leq \expt{S}-Nx] \leq \exp\left( -\frac{N^2x^2}{2\sum_{i=1}^{N}\expt{X_i^2}}\right).
\end{equation}
\end{theorem}

In our case where $\gamma_{j,i}$s are i.i.d. with finite means $\mu$ and variances $\sigma^2$, and $Z= \sum_{\substack{ j\neq i  \\ j\in \set{S}}}%
\gamma _{j,i}$ with $\abs{\set{S}}=m$, we have $\expt{Z}=(m-1)\mu$ and $\expt{\gamma_{j,i}^2}=\mu^2+\sigma^2$. Then, the above inequality can be written as
\begin{equation}
\Pr[Z\leq (m-1)(\mu-x)] \leq \exp\left( -\frac{(m-1)x^2}{2(\mu^2+\sigma^2)}\right).
\end{equation}
Finally, in order to have the form of $\Pr[Z\leq s]$ as in \eqref{eq:each_one_tbb}, set $(m-1)(\mu-x)=s$ and accordingly we have
\begin{equation}
\label{eq:Pr_Z_s}
\Pr[Z\leq s] \leq \exp\left( -\frac{(m-1)(\mu-\frac{s}{m-1})^2}{2(\mu^2+\sigma^2)}\right),
\end{equation}
for $s<(m-1)\mu$ (due to the fact $x>0$).

Next, we upper-bound $\Pr \bigl[ \frac{\gamma _{i,i}}{1/\rho +s}\geq \beta _{0}\bigr]$:
\begin{eqnarray}
\lefteqn{\Pr \biggl[ \frac{\gamma_{i,i}}{1/\rho +s}\geq \beta _{0}\biggr]} \notag \\
&&=
\Pr\bigl[ \gamma_{i,i} \geq \beta_{0}(1/\rho +s)\bigr] \notag \\
&& = \Pr\bigl[ \gamma_{i,i}-\mu \geq \beta_{0}(1/\rho +s)-\mu \bigr] \notag \\
&& \leq \Pr\bigl[ \abs{\gamma_{i,i}-\mu} \geq \beta_{0}(1/\rho +s)-\mu \bigr] \label{eq:RHS_gep_zero} \\
&& \leq \frac{\sigma^2}{(\beta_0s+\beta_0/\rho-\mu)^2}, \label{eq:chebyshev}
\end{eqnarray}
where \eqref{eq:chebyshev} follows from the Chebyshev inequality. Note that for \eqref{eq:RHS_gep_zero} to hold, we need $\beta_{0}(1/\rho +s)-\mu\geq 0$, i.e., $s\geq \frac{\mu}{\beta_0}-\frac{1}{\rho}$.

Substituting \eqref{eq:Pr_Z_s} and \eqref{eq:chebyshev} into \eqref{eq:each_one_tbb}, with $\frac{\mu}{\beta_0}-\frac{1}{\rho}\leq s<(m-1)\mu$, we have
\begin{equation}
\label{eq:final_Pr_SINR_UB}
\begin{split}
\Pr \left[ \frac{\gamma_{i,i}}{1/\rho +Z} \geq \beta _{0}\right]
&\leq \exp\left( -\frac{(m-1)(\mu-\frac{s}{m-1})^2}{2(\mu^2+\sigma^2)}\right) \\
&\quad +\frac{\sigma^2}{(\beta_0s+\beta_0/\rho-\mu)^2}.\\
\end{split}
\end{equation}

Note that \eqref{eq:final_Pr_SINR_UB} holds for all $s$ satisfying $\frac{\mu}{\beta_0}-\frac{1}{\rho}\leq s<(m-1)\mu$. In the problem under discussion, it suffices to set $s=\frac{m-1}{2}\mu$.\footnote{Here, we implicitly assume $s=\frac{m-1}{2}\mu\geq \frac{\mu}{\beta_0}-\frac{1}{\rho}$, which holds readily as $s=\frac{m-1}{2}\mu$ scales with $m$ whereas $\frac{\mu}{\beta_0}-\frac{1}{\rho}$ is a constant.} Then it follows immediately that the first term on the RHS of \eqref{eq:final_Pr_SINR_UB} decays exponentially with $m$, while the second term decays as $\frac{1}{m^2}$. Thus, as far as scaling is concerned, $\Pr\Bigl[\frac{\gamma_{i,i}}{\frac{1}{\rho}+\sum
_{j\neq i,\,j\in \set{S}}\gamma_{j,i}}\geq \beta _{0}\Bigr]$ is upper-bounded
by $O(1/m^2)$.

\section{Derivation of \eqref{eq:Pr_Mn_geq_EMn}}
\label{app:Pr_Mn_geq_EMn}

From \eqref{eq:CDF}, the cdf of the extreme value $M_{n}$ can be
written as follows (with the assumption of $\mu =\sigma =1$)
\begin{equation}
F_{M_{n}}(x)=\left( F_{n}(x)\right) ^{n}=\biggl(\frac{x-1}{\sqrt{2n-1%
}}\frac{n-1}{n}+\frac{1}{n}\biggr)^{n/(n-1)}.  \label{eq:CDF_Mn}
\end{equation}%
It can be easily verified that $\lim_{n\rightarrow \infty }$ $F_{M_n}(%
\expt{M_n})=1/2$. For increasing $n,$ we have $\mathbb{E}\left[
M_{n}\right] \approx \frac{n-1}{\sqrt{2n-1}}$; it follows that
\begin{equation*}
\PP\biggl[M_{n}>\frac{n-1}{\sqrt{2n-1}}\biggr]=\PP\biggl[M_{n}\leq \frac{n-1%
}{\sqrt{2n-1}}\biggr]=\frac{1}{2}.
\end{equation*}

\section{Proof of Lemma~\protect\ref{lem:Pr_Mn_In_geq_0}}

\label{app:prove_Pr_Mn_In_geq_0}

We start by quoting the following results from \cite[Problem~26,
pp.~465]{Feller_vol2:71}. (The proof follows readily from the hint
for the solution, and therefore it is omitted here.) If the
distribution satisfies the conditions of Lemma~\ref
{lem:Pr_Mn_In_geq_0}, it is known that:

\begin{enumerate}
\item The distribution of the ratio $I_{n}/M_{n}$ has a Laplace transform $%
w_{n}(\lambda )$ converging to $w(\lambda )=\frac{e^{-\lambda
}}{1+\nu \int_{0}^{1}(1-e^{-\lambda t})t^{-\nu -1}\ud t}$;

\item $\expt{I_n/M_n}\rightarrow 1/(1-\nu)$.
\end{enumerate}

Now, we can proceed with proving $\Pr [{M_{n}}/{I_{n}}\geq 1-\nu
]>0$ by contradiction. Suppose $\Pr [{M_{n}}/{I_{n}}\geq 1-\nu ]=0$,
then
\begin{align}
\expt{M_{n}/I_{n}}& =\int_{0}^{\infty }(1-F(x))\,\ud x \label{eq:E_for_cts_rv}\\
& =\int_{0}^{\infty }\Pr [{M_{n}}/{I_{n}}\geq x]\,\ud x \\
& =\int_{0}^{1-\nu }\Pr [{M_{n}}/{I_{n}}\geq x]\,\ud x \notag \\
&\quad +\underbrace{%
\int_{1-\nu }^{\infty }\Pr [{M_{n}}/{I_{n}}\geq x]\,\ud x}_{=0} \\
& \leq 1-\nu,  \label{eq:EX_4}
\end{align}
where \eqref{eq:E_for_cts_rv} follows from the fact that $\expt{X}=\int_{0}^{\infty}(1-F(x))\ud x$ for continuous and nonnegative random variables.

However, by Jensen's inequality, we have $\expt{M_n/I_n}\geq
1/\expt{I_n/M_n} =1-\nu$. Therefore, we have $\expt{M_n/I_n}=1-\nu$,
implying that equality holds in Jensen's inequality. This is only
when the distribution of the random variable $I_{n}/M_{n}$ is
degenerate (i.e., $I_{n}/M_{n}$ is a constant). However, this
contradicts property $1$) which implies a non-degenerate
distribution.

Thus, $\Pr [{M_{n}}/{I_{n}}\geq 1-\nu ]>0$ must hold.

\section{Average Number of Scheduled Transmissions is $\Theta(n)$}

\label{app:show_Theta_n}

Under the setting of $m=n$ (i.e., deploying as many relay nodes as
S--D pairs), it is shown that the average number of scheduled
concurrent transmissions in the two-hop opportunistic relaying
scheme is of the order of $n$. To see this, define
\begin{equation*}
Z_{i}=%
\begin{cases}
1, & \text{if receiver $i$ is scheduled,} \\
0, & \text{otherwise,}%
\end{cases}%
\end{equation*}%
so that the number of scheduled sources is written as
\begin{equation*}
Z=Z_{1}+\cdots +Z_{n}.
\end{equation*}%
Linearity of the expectation yields,
\begin{equation*}
\expt{Z}=\expt{Z_{1}}+\cdots +\expt{Z_{n}}.
\end{equation*}%
For each $i$, $Z_{i}$ has a simple Bernoulli distribution. Specifically, in
the i.i.d. random model of Section~\ref{sec:sysmod}, each source has
probability of $1/n$ to be selected by any relay. We have
\begin{align*}
\expt{Z_{i}}& =\Pr [\text{node $i$ is scheduled}] \\
& =1-\Pr [\text{node $i$ is not scheduled}] \\
& =1-\biggl(1-\frac{1}{n}\biggr)^{n} \\
& \rightarrow 1-e^{-1},
\end{align*}%
where the last equation follows from the fact $(1-x)^{n}\rightarrow e^{-nx}$
for $x\rightarrow 0$. Finally, we readily have
\begin{equation}
\expt{Z}=n(1-e^{-1}),  \label{eq:avg_sch_trans}
\end{equation}%
which is $\Theta (n)$.

\bibliographystyle{IEEEtran}
\bibliography{throughput_scaling_final,IEEEabrv}

\end{document}